\documentclass[aps,prd,twocolumn,groupedaddress]{revtex4-1}
\usepackage{graphicx}
\usepackage{wrapfig}
\usepackage{latexsym}
\usepackage{amssymb}
\usepackage{amsfonts,color}
\usepackage{float}
\usepackage{amsmath}

\newcommand{\be}{\begin{equation}}
\newcommand{\ee}{\end{equation}}
\newcommand{\ba}{\begin{align}}
\newcommand{\ea}{\end{align}}

\begin{document}


\title{The Radial Acceleration Relation and a Magnetostatic Analogy in Quasilinear MOND}
\author{Katherine Brown$^{(1)}$, Roshan Abraham$^{(2)}$, Leo Kell$^{(1)}$, and Harsh Mathur$^{(2)}$} 
\affiliation{$^{(1)}$Department of Physics, Hamilton College, Clinton, NY 13323}

\vspace{1mm}
\affiliation{$^{(2)}$Department of Physics, Case Western Reserve University, Cleveland, Ohio 44106-7079}


\date{\today}

\begin{abstract}
Recently a remarkable relation has been demonstrated between the observed radial acceleration in disk galaxies and the 
acceleration predicted on the basis of baryonic matter alone. Here we study this relation within the framework of the modified 
gravity model MOND. The field equations of MOND automatically imply the radial acceleration relation for spherically 
symmetric galaxies, but for disk galaxies deviations from the relation are expected. Here we investigate whether these 
deviations are of sufficient magnitude to bring MOND into conflict with the observed relation. In the quasilinear formulation of 
MOND, to calculate the gravitational field of a given distribution of matter, an intermediate step is to calculate the ``pristine 
field'', which is a simple nonlinear function of the Newtonian field corresponding to the same distribution of matter. Hence, to 
the extent that the quasilinear gravitational field is approximately equal to the pristine field, the radial acceleration relation will 
be satisfied. We show that the difference between the quasilinear and pristine fields obeys the equations of magnetostatics; 
the curl of the pristine field serves as the source for the difference in the two fields, much as currents serve as sources for the 
magnetic field. Using the magnetostatic analogy we numerically study the difference between the pristine and quasilinear 
fields for simple model galaxies with a Gaussian profile. Our principal finding is that the difference between the fields is small 
compared to the observational uncertainties and that quasilinear MOND is therefore compatible with the observed radial 
acceleration relation.

\end{abstract}

\pacs{}

\maketitle

\section{Introduction \label{introduction}}
The problem of dark matter is one of the central problems in
modern cosmology \cite{weinberg}. According to the conventional $\Lambda$CDM
paradigm, matter constitutes less than a third of the
total energy density of the Universe. The bulk of this matter is non-baryonic. 
Many lines of evidence point to the existence of dark matter, notably
the cosmic microwave background anisotropies \cite{planck} and observations of
lensing \cite{lens}, but dark matter has eluded direct detection \cite{dan} and its nature
remains a mystery.

Rotation curves of disk galaxies provided the earliest compelling
evidence for dark matter \cite{rubin}. The observation that the rotation velocity
generically approaches a plateau instead of vanishing asymptotically
is conventionally interpreted as evidence for a spherical dark
matter halo that dominates the visible baryonic
matter in the disk. Puzzling from this perspective were empirical
observations such as the baryonic Tully-Fisher relation \cite{tully} that appear to
suggest a strong connection between the rotation curves and the 
distribution of baryonic matter. 

In an important recent study McGaugh {\em et al.} \cite{stacy1, stacy2} demonstrated
a relation between the observed radial acceleration in disk galaxies
and the acceleration predicted on the basis of baryonic matter alone.  
This empirical relation, dubbed the ``radial acceleration relation'' (RAR)
is based on an impressive body of data, the SPARC dataset that represents
near infrared observations by the Spitzer Space Telescope and 
21 cm observations by radio astronomers over several decades \cite{stacy3}. The radial acceleration relation 
has to be explained
by any tenable model of dark matter. There have already been several
efforts to show that this relation is compatible with $\Lambda$CDM
cosmology\cite{keller, ludlow, kosowsky}. There have also been efforts to understand this relationship
using modified theories of gravity ranging from scalar tensor theories \cite{burrage}
to emergent gravity models \cite{niz, emergent}. 

The modified gravity theory MOND was originally introduced \cite{milgrom1} to account
for galaxy rotation and it has had some phenomenological success,
e.g., providing an explanation for the baryonic Tully-Fisher relation.
However MOND and its relativistic completions have well-known problems on 
cluster and cosmological scales \cite{glenn, mondreview, bullet}. 
It is sometimes assumed that MOND predicts
the observed radial acceleration relation, but in fact the MOND field
equations imply such a relationship only for spherically-symmetric galaxies \cite{milgrombrada}.
For disk galaxies deviations are to be expected. 
With the empirical discovery of the remarkable radial acceleration relation
it becomes imperative to determine the magnitude of these deviations;
that is the object of this
paper. Our principle finding is that the deviations are 
small compared to the experimental uncertainties and that the
MOND field equations are in fact compatible with the observed radial
acceleration relation.

Our analysis is carried out in the framework of `quasilinear' MOND \cite{quasi}. 
Notwithstanding the name, quasilinear MOND is a highly nonlinear
theory. In order to calculate 
the quasilinear gravitational field, ${\mathbf g}_Q$, corresponding to a distribution of matter $\rho_D$,
an intermediate step is to calculate a ``pristine field'', ${\mathbf g}_P$. The pristine field is related to the Newtonian field through a simple, nonlinear `interpolating function', thus, if one was to plot 
observed versus Newtonian accelerations within pristine MOND, finding agreement with the radial acceleration relation would simply be a matter of 
choosing the right interpolating function \cite{stacy2}.
To the extent that the quasilinear field is approximately
equal to the pristine field, it will obey the radial acceleration
relation. In order to demonstrate the approximate equality of the two fields we
show that the difference ${\mathbf b} = {\mathbf g}_P - {\mathbf g}_Q$
obeys the equations of magnetostatics with ${\mathbf b}$ analogous
to the magnetic field and $\nabla \times {\mathbf g}_P$ analogous to
the current that sources ${\mathbf b}$. This analogy provides
useful qualitative insights into the deviation field ${\mathbf b}$
as well as a practical numerical scheme for its calculation.

Although the main focus of this paper is on modified gravity formulations of MOND it is worth noting that there is another class of models called modified inertia formulations of MOND. Although modified inertia models have received comparatively less attention \cite{milgrom2}, there has been an attempt to construct a relativistic formulation \cite{demir} and an interesting experimental test has been proposed \cite{ignatiev}. It would be desirable in future work to explore possible contrasts between the predictions of modified inertia and modified gravity theories of MOND for galactic motion along the lines suggested by \cite{milgrom2}.

For simplicity we study simple model galaxies with a Gaussian profile.
Our methods as well as our results will easily carry over to other more
realistic galaxy models. The numerical analysis is most efficiently done
by use of the fast Fourier transform although there are some subtleties
that arise due to the long range nature of gravitational forces.
Key results are presented in fig \ref{fig:helmholtz} which 
shows that the current distribution $\nabla \times {\mathbf g}_P$ for a disk galaxy 
resembles a Helmholtz gradient coil; fig \ref{fig:b} in which the dependence
of the deviation ${\mathbf b}$ on distance from the galactic center
and on the aspect ratio of the disk galaxy is systematically investigated;
and fig \ref{fig:streaky} in which we display the computed radial
acceleration relation for our model galaxies. 
Comparison of this plot with the observed
radial acceleration relation leads to our conclusion. 

\section{Magnetostatic analogy for MOND gravity}
\label{analogy}

\subsection{Quasilinear MOND}
\label{sec:mond}

In Newtonian gravity, the acceleration due to gravity ${\mathbf g}_N$ obeys the field equations
\begin{equation}
\nabla \cdot {\mathbf g}_N = - 4 \pi G \rho ({\mathbf r}) 
\hspace{3mm} {\rm and} \hspace{3mm} \nabla \times {\mathbf g}_N = 0.
\label{eq:newton}
\end{equation}
Here $\rho ({\mathbf r})$ is the mass distribution producing the gravitational field. 
In the pristine formulation of MOND \cite{milgrom1} a test particle however experiences an acceleration 
${\mathbf g}_P$ which is related to the Newtonian acceleration via the algebraic relation 
\begin{equation}
{\mathbf g}_P = f \left( \frac{ g_N }{a_0} \right) {\mathbf g}_N 
\label{eq:pristine}
\end{equation}
Here $a_0 \sim 10^{-10}$ m/s$^2$ is the MOND acceleration scale and the interpolating
function $f$ has the asymptotic behavior $f(x) \approx 1$ for $x \gg 1$ and $f(x) \approx x^{-1/2}$
for $x \ll 1$. Thus the observed acceleration of the test particle matches the Newtonian prediction when the
Newtonian gravitational field is strong but is significantly enhanced when the Newtonian field is weak on the
MOND scale. 
Pristine MOND exactly reproduces the radial acceleration relation; 
indeed the observed radial acceleration relation  may be regarded as a measurement of
the MOND acceleration $a_0$ and the interpolating function $f$. 

However, pristine MOND suffers from physical inconsistencies and lacks a Lagrangian formulation. 
Thus it cannot be considered a viable alternative theory of gravitation as noted in \cite{quasi} and \cite{bekenstein}. 
 These authors introduced nonlinear MOND \cite{bekenstein} and quasilinear MOND \cite{quasi} 
as viable alternatives to Newtonian gravity; in this paper we focus on quasilinear MOND.  
Within quasilinear MOND the acceleration of the test particle is
not equal to the pristine field ${\mathbf g}_P$ but to the quasilinear field 
${\mathbf g}_Q$ which is determined by solving

\begin{equation}
\nabla \cdot {\mathbf g}_Q = \nabla \cdot {\mathbf g}_P \equiv - 4 \pi G \rho_{{\rm eq}} 
\hspace{3mm} {\rm and} \hspace{3mm}
\nabla \times {\mathbf g}_Q = 0.
\label{eq:quasilinear}
\end{equation}
Invoking the Helmholtz decomposition of vector fields we see that ${\mathbf g}_Q$ is the curl 
free part of ${\mathbf g}_P$. The divergence of ${\mathbf g}_P$ (or more precisely $\rho_{{\rm eq}}$) has
the interpretation of being the combined density of dark and baryonic matter that would be required to produce the
same observed acceleration of the test mass if Newtonian gravity prevailed.

To complete the formulation of quasilinear MOND we must specify the interpolating function
$f$. A simple form that has the desired asymptotic behavior is
\begin{equation}
f(x) = \sqrt{ 1 + \frac{1}{x} }.
\label{eq:sqrt}
\end{equation}
An alternative form, following ref \cite{stacy1}, is to take 
\begin{equation}
f(x) = \frac{1}{ 1 - \exp ( - \sqrt{x} ) }.
\label{eq:planck}
\end{equation}

\subsection{Magnetostatic Analogy}
\label{sec:magnetostatic}

For a spherically symmetric galaxy ${\mathbf g}_P$ is curl free and hence ${\mathbf g}_Q = {\mathbf g}_P$.
In this case it automatically follows that the observed acceleration ${\mathbf g}_Q$ will be a simple function of
the Newtonian gravitational field produced by the baryonic matter in the galaxy, namely, 
${\mathbf g}_Q = {\mathbf g}_N f(g_N/a_0)$. However only in cases of exceptional symmetry 
will ${\mathbf g}_P$ be curl free. Generically ${\mathbf g}_Q$ is not equal to ${\mathbf g}_P$
and hence ${\mathbf g}_Q$ is not a local function of ${\mathbf g}_N$. For the radial acceleration relation to
hold in quasilinear MOND we need ${\mathbf g}_Q \approx {\mathbf g}_P$ at least in the galactic plane. 

Whether the approximate equality ${\mathbf g}_Q \approx {\mathbf g}_P$ holds for disc galaxies is the key question that is
studied in this paper. In \cite{milgrombrada}  the corresponding problem was pointed out for nonlinear MOND and an
estimate was made of the size of the discrepancy between the observed acceleration and the pristine
field strength in that theory. The new high quality data establishing the radial acceleration relation
provides the impetus for the detailed study of quasilinear MOND reported in this paper.

To this end it is convenient to define 
\be
{\mathbf b} = {\mathbf g}_P - {\mathbf g}_Q
\label{eq:bdefn}
\end{equation}
and 
\begin{equation}
{\mathbf j} = \nabla \times {\mathbf g}_P.
\label{eq:jdefn}
\end{equation}
It then follows from eq (\ref{eq:quasilinear}) that
\begin{equation}
\nabla \cdot {\mathbf b} = 0 \hspace{3mm}
{\rm and} \hspace{3mm}
\nabla \times {\mathbf b} = {\mathbf j}.
\label{eq:magnetostatics}
\end{equation}
These are precisely the equations of magnetostatics. 
We have already noted that $\nabla \cdot {\mathbf g}_P$ may be interpreted as the equivalent density of
dark matter needed to reproduce MOND behavior within Newtonian gravity. 
Eq (\ref{eq:magnetostatics}) reveals that $\nabla \times {\mathbf g}_P$ may be interpreted as a
current that sources the difference between
${\mathbf g}_P$  and ${\mathbf g}_Q$. 

Given ${\mathbf g}_P$ eq (\ref{eq:magnetostatics}) provides an efficient way to
calculate ${\mathbf b}$ and thereby ${\mathbf g_Q}$ as we will discuss below.
Formally one can write down a solution to eq (\ref{eq:magnetostatics}) expressing
${\mathbf b}$ in terms 
of ${\mathbf j}$ using the Biot-Savart law. Whereas the relation eq (\ref{eq:pristine}) is 
local in the sense that ${\mathbf g}_P$ at a given point is determined by 
  ${\mathbf g}_N$ at the same point, in sharp contrast the relation between ${\mathbf b}$ and
  ${\mathbf j}$ is not local. This shows that in general ${\mathbf g}_Q$ is not a local
  function of ${\mathbf g}_N$ in quasilinear MOND.

\section{Simulated Radial Acceleration Relations for Model Galaxies}

\label{sec:rar}

\subsection{Model and Methods}

\label{sec:model}

We now introduce a simple model which allows us to study the deviation from 
the radial acceleration relation in disk galaxies. We take the galaxies to have a Gaussian profile
\be
\rho_D = \frac{M}{\pi^{3/2}\Delta R_d^3} \exp\left[-\frac{(x^2 + y ^2)}{R_d^2} \right] \exp \left[ - \frac{z^2}{\Delta^2 R_d^2} \right].
\label{eq:profile}
\ee
Here $M$ is the mass of the galaxy and $R_d$ is its radius. $\Delta$ is a dimensionless parameter that characterizes
the aspect ratio of the galaxy; $\Delta = 1$ corresponds to a spherically symmetric galaxy and $\Delta \rightarrow 0$
to a flat disk. The characteristic radial
acceleration scale within Newtonian gravity for this circumstance is $GM/R_d^2$. The 
MOND field of the galaxy is therefore fully determined by the dimensionless parameters $\mu = GM/R_d^2 a_0$ and $\Delta$. 
We will work in a system of units wherein $R_d = 1$ and $a_0 = 1$. Then the parameter $\mu = GM$ 
corresponds to a measure of the mass of the galaxy in our system of units. 
Other simple models used to describe
disk galaxies \cite{binney} can be treated using our methods and will lead to similar results; 
the advantage of using
the Gaussian model is that it allows us to do higher resolution
numerics and to rapidly explore a broader range of MOND parameters. 
Hence we focus on 
the Gaussian model here.

The first step in our analysis is to calculate ${\mathbf g}_N$, the Newtonian field of the galaxy. Note that
it is sufficient to calculate ${\mathbf g}_N$ for $\mu =1$ since by linearity ${\mathbf g}_N \propto \mu$. 
Eq (\ref{eq:newton}) has the simple solution
\begin{equation}
\tilde{\mathbf g}_N ({\mathbf k}) = i 4 \pi G \; {\mathbf k} \frac{1}{k^2}  \tilde{\rho}_D ({\mathbf k}) 
\label{eq:fourier}
\end{equation}
in Fourier space. (A tilde above a symbol denotes the Fourier transform.)
The Newtonian field can therefore be efficiently calculated using the fast Fourier transform.
We assume that the galaxy is located at the center of a cube of side $L$ and we
sample functions at points on a regular cubic lattice with $2N+1$ points per edge of the cube. 

An amusing subtlety arises because of the long-range nature of the gravitational force. By using
the Fourier transform we enforce a periodicity in the mass distribution and the gravitational
field whereas in the real problem we wish to impose the boundary condition that the gravitational
field vanishes far away from the galaxy. Naively one might suppose that in the limit that $L$ is sufficiently
large the results would be insensitive to boundary conditions but it is easy to demonstrate that in 
fact a sensitivity to boundary conditions persists in the limit $L \rightarrow \infty$ because of the
inverse square falloff of the gravitational field \cite{olbers}.
The remedy is to calculate not the gravitational field
of the disk galaxy directly but rather of the mass distribution $\rho_D - \rho_{{\rm sph}}$. Here
$\rho_{{\rm sph}}$ is a spherically symmetric distribution with a total mass, $M$, the same as
the galaxy distribution $\rho_D$. The advantage is that the gravitational field of the difference
falls off like a quadrupole as $1/r^4$. This is a sufficiently rapid falloff that the result is insensitive
to boundary conditions for sufficiently large $L$. We can then construct the gravitational field
corresponding to the galaxy distribution $\rho_G$ by adding the field corresponding to the
spherical distribution $\rho_{{\rm sph}}$ which can be obtained analytically.
In practice we take $\rho_{{\rm sph}}$ to be an isotropic Gaussian with radius $R_d \Delta^{1/3}$. 

Calculation of ${\mathbf g}_P$ via eq (\ref{eq:quasilinear}) is now a simple matter of applying an algebraic function
to ${\mathbf g}_N$.
${\mathbf b}$ can be calculated by solving eq (\ref{eq:magnetostatics}) by Fourier methods. In order to 
calculate the curl of ${\mathbf g}_P$ it is helpful to first subtract a suitable isotropic function to ensure 
that the results are insensitive to boundary conditions much as we do in our computation of
${\mathbf g}_N$. In practice we subtract the pristine field corresponding
to the isotropic distribution $\rho_{{\rm sph}}$ to facilitate the computations. 
Once ${\mathbf g}_P$ and ${\mathbf b}$ are determined it is a simple matter
to obtain ${\mathbf g}_Q$ using eq (\ref{eq:bdefn}). 

In the results presented here we consider aspect ratios in the range $0.5 < \Delta < 1.0$ 
and the mass parameter in the range $0.01 < \mu < 10.0$. Over this range of parameters 
it is sufficient to calculate the Newtonian field using $L = 4$ and $N = 36$. Smaller values of $\Delta$ are more realistic for disk galaxies but
would necessitate a finer grid. However in order to show the qualitative
difference between pristine and quasilinear MOND, and to show that quasilinear MOND reproduces the
observed radial acceleration relation it is sufficient to consider $\Delta = 0.5$ as discussed further in 
section \ref{sec:results}; we have verified that the results extrapolate smoothly with $\Delta$ down to 
$\Delta = 0.1$. 
The results are also
insensitive to variations the size of the box and to the fineness of the discretization. 
We have also checked the accuracy of the calculated Newtonian field by using the 
multipole expansion to calculate the large distance asymptotic Newtonian field
corresponding to the density profile eq (\ref{eq:profile}). To check the accuracy of
the calculated Newtonian field at short distances we have
performed a non-trivial 
check described in appendix \ref{sec:appendixa}. 
The subsequent calculation of
${\mathbf g}_P$ is limited only by numerical roundoff since it only involves solving an
algebraic relation. The calculation of ${\mathbf g}_Q$ from ${\mathbf g}_P$ is 
similar to the calculation of ${\mathbf g}_N$ and can be checked in similar ways.

\subsection{Results}

\label{sec:results}

\begin{figure}[h]
\includegraphics[width=3.5in]{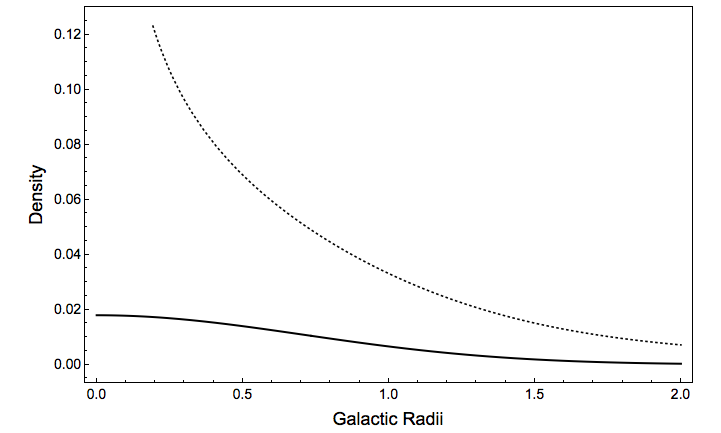}
\caption{Dimensionless quantities $G \rho_D R_d/ a_0$ (solid) and $G \rho_{{\rm eq}} R_d/a_0$ (dotted) showing the radial density profile for an isotropic galaxy with $\Delta=1, \mu = 0.1$.}
\label{fig:eqdensity}
\end{figure}

We start with a discussion of the quantity $\rho_{{\rm eq}}$ 
introduced in eq (\ref{eq:quasilinear}).
If we cast a dark matter interpretation upon MOND we may regard $\rho_D$ as the density of baryonic
matter and $\rho_{{\rm eq}}$ as the combined density of baryonic and dark matter. 
Fig \ref{fig:eqdensity}
shows the two 
densities for an isotropic galaxy with aspect ratio $\Delta = 1$ and mass parameter $\mu = 0.1$. 
The plot of $\rho_D$ in the left panel corresponds to the Gaussian profile eq (\ref{eq:profile}). 
The plot of $\rho_{{\rm eq}}$ in the right panel shows three interesting features. First the density 
$\rho_{{\rm eq}}$ is much greater than $\rho_D$, 
consistent with the preponderance of dark matter compared to baryonic matter in the 
conventional $\Lambda$CDM paradigm. Second the density extends to a greater distance
from the galactic center much like a dark matter halo in the conventional $\Lambda$CDM picture.
Third the density has a weak divergence near the center of the galaxy ($\rho_{{\rm eq}} \propto 1/\sqrt{r}$ 
for $r \rightarrow 0$); this is a generic feature of quasilinear MOND and it should be noted that it does not
imply that the core of the galaxy has an infinite mass since the divergence is integrable. 
The plots in fig \ref{fig:eqdensity} were generated using analytic expressions for ${\mathbf g}_N$ and
${\mathbf g}_P$ which are available in the isotropic case. Our numerical methods can be employed
to examine the behavior of $\rho_{{\rm eq}}$ for disk galaxies. Those results are of interest but they
will be reported elsewhere \cite{scott} since they are tangential to the present study. 

\begin{figure}
\includegraphics[width=3.5in]{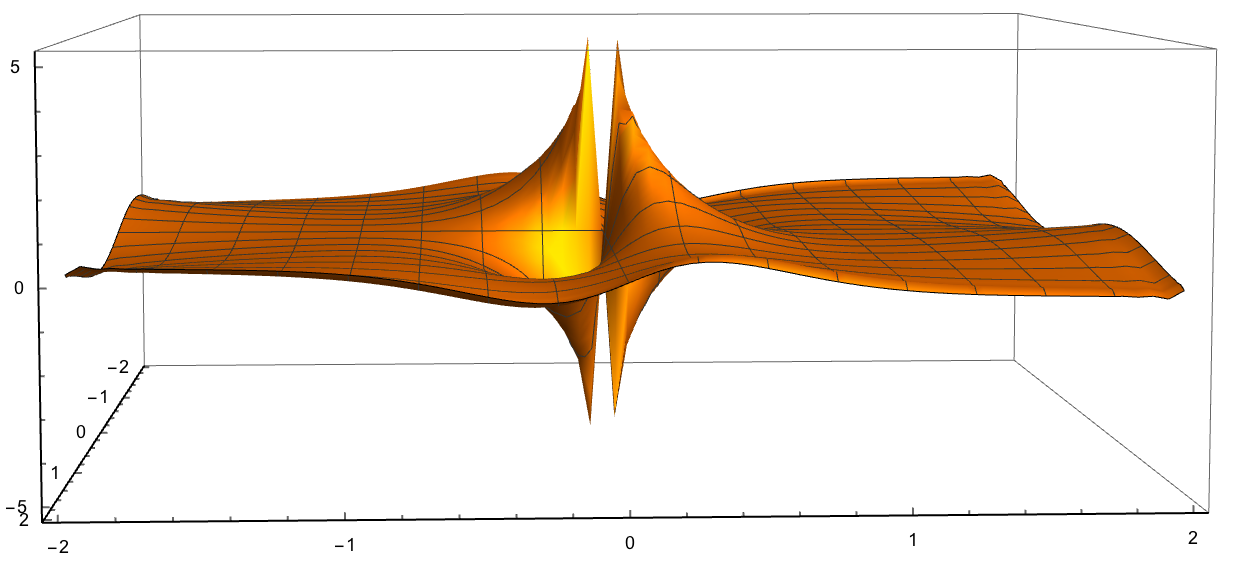}
\caption{The $y$ component of ${\mathbf j} = \nabla \times
{\mathbf g}_P$ vs position in the $x$-$z$ plane for a galaxy with $\Delta = 0.5$ and $\mu = 0.1$. 
By symmetry the $x$ and
$z$ components should vanish and this is verified by
by the numerics.} 
\label{fig:helmholtz}
\end{figure}

According to the magnetostatic analogy, the quantity ${\mathbf j} = \nabla \times {\mathbf g}_P$ may be 
regarded as a current that serves as the source for ${\mathbf b}$, the difference between the pristine
and quasilinear MOND fields. Using the product rule it is easy to see that $\nabla \times {\mathbf g}_P = 
f'(g_N/a_0) \; {\mathbf g}_N \times \nabla (g_N/a_0)$. Due to the azimuthal symmetry of the galaxy
it is easy to see that neither ${\mathbf g}_N$ nor $\nabla g_N$ can have an azimuthal component; it 
follows that ${\mathbf j}$ must be purely azimuthal. Thus symmetry requires that 
the current that sources ${\mathbf b}$ circulates about the axis of the galaxy. Similarly
the assumed reflection symmetry of the galaxy about the $x$-$y$ plane implies that the 
current ${\mathbf j}$ must be antisymmetric under reflection about the $x$-$y$ plane. 
Thus the currents circulate in opposite senses above and below the galactic plane. This is as 
far as symmetry arguments go. Fig \ref{fig:helmholtz} shows an explicit numerical calculation 
of the $y$ component of the current as a function of position in the $x$-$y$ plane
for a galaxy with aspect ratio $\Delta = 0.5$ and mass parameter $\mu = 0.1$. 
The plot shows two sharp peaks and two sharp anti-peaks that are related by the
azimuthal and reflection symmetries noted above. It follows that the current is concentrated
into two counter-propagating circular coils that are parallel to the $x$-$y$ plane, one located
above the plane and the other below it, analogous to a Helmholtz gradient coil. We can easily
draw upon electromagnetic intuition to draw the lines of ${\mathbf b}$ or to recognize that
at large distance the resulting field will be quadrupolar.

\begin{figure*}
\includegraphics[width=\textwidth]{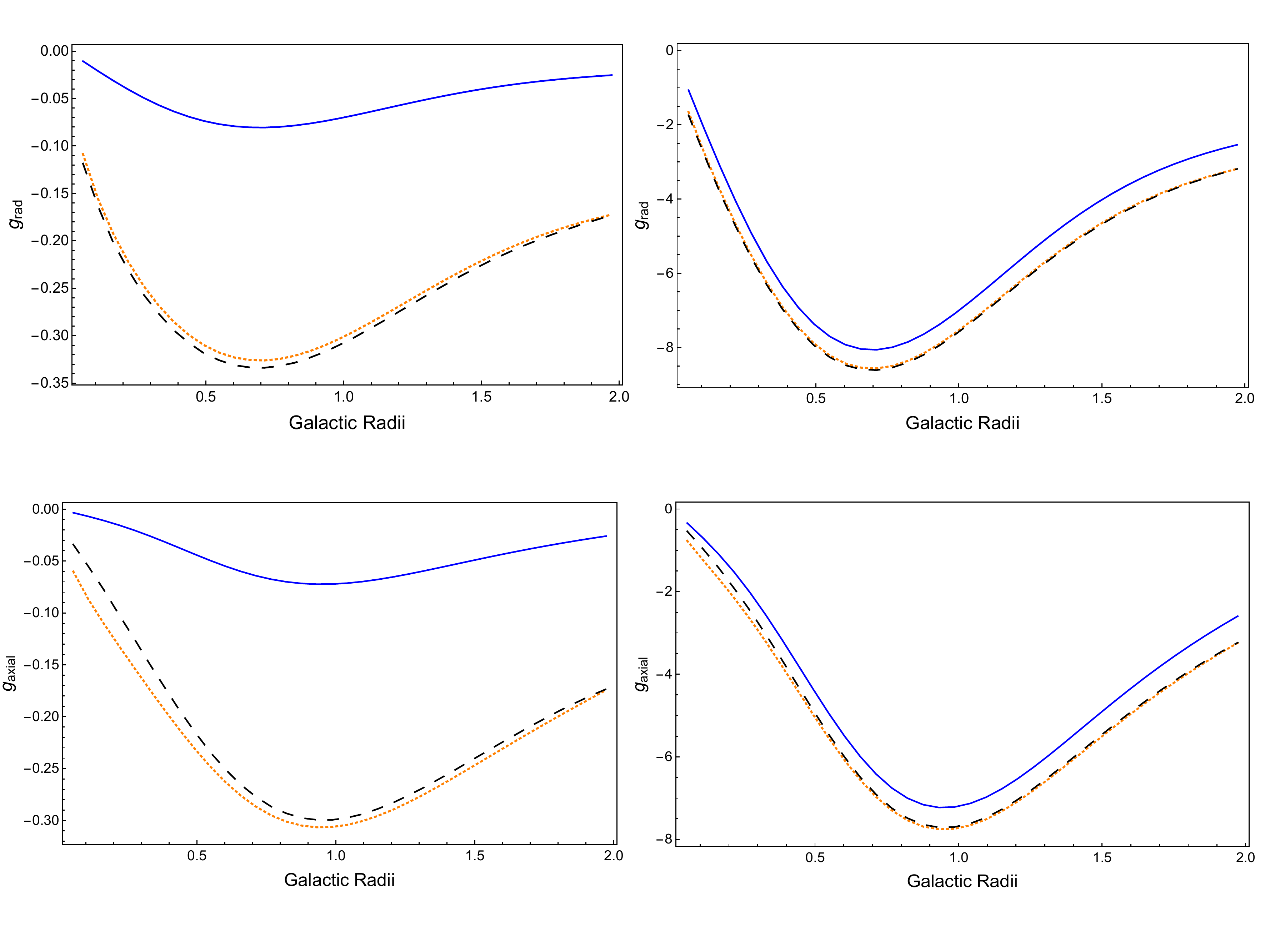}
\caption{Newtonian (solid blue), pristine (solid orange) and quasilinear (dashed) gravitational fields for a galaxy with $\Delta = 0.5$. The upper panels show the radial component plotted vs $r$; the lower panels, the axial component vs $z$.  Deep MOND regime ($\mu = 0.1$) on the left and Newtonian regime ($\mu = 10$) on the right in both panels. 
Note the much larger deviation of 
$g_P$ and $g_Q$ from $g_N$ for the smaller values of $\mu$. Note also that the deviation between
$g_P$ and $g_Q$ is more pronounced in the axial plots than the radial plots; the sign of the 
deviation is also opposite in the two cases.}
\label{fig:radialaxial}
\end{figure*}

We turn now to a comparison of the Newtonian, pristine MOND and quasilinear MOND gravitational
fields. The upper panels of Fig.\ref{fig:radialaxial} plot the radial component of the field as a function of the cylindrical coordinate $r$. Each field is computed at 36 points in the galactic
plane at distances that vary uniformly from zero to two galactic radii. 
Both plots are 
for galaxies with aspect ratio $\Delta = 0.5$. The plot at left is for a galaxy that is deep in the MOND regime
with parameter $\mu = 0.1$; the plot on the right is deep in the Newtonian regime with parameter $\mu = 10.$
As expected the Newtonian field shown in blue vanishes both as $r \rightarrow 0$ and $r \rightarrow \infty$. 
We can check the large $r$ behavior using a multipole expansion and we can also perform a non-trivial 
check at small $r$ as described in Appendix \ref{sec:appendixa}. 
The orange curves correspond to the
pristine MOND acceleration. This is simply related to the Newtonian field via the algebraic function
eq (\ref{eq:pristine}). The plots show that for small $\mu$ the pristine 
field is considerably
enhanced relative to the Newtonian field but for large $\mu$ it is essentially equal to the Newtonian field. 
In fig \ref{fig:radialaxial} we have used the interpolating function in eq (\ref{eq:sqrt}) 
but we have explicitly verified that the alternative interpolation eq (\ref{eq:planck}) gives similar results. 
The black dashed curves show the radial component of the quasilinear 
acceleration. 
For this aspect ratio it appears that there is very little difference between the pristine 
and quasilinear 
fields at least in the galactic plane. We will examine this difference more
closely below. The lower two panels of Fig.\ref{fig:radialaxial} show the $z$ component of the Newtonian, pristine, and quasilinear fields as a function of $z$. A key difference from the radial plots is that in this case the
 magnitude of the pristine 
 acceleration is bigger than the magnitude of the quasilinear 
 acceleration. This difference can be understood by picturing the difference between the
 two fields, ${\mathbf b}$, as the magnetic field of a Helmholtz gradient coil. Another noteworthy
 feature is that the difference between the pristine and quasilinear fields is bigger along the
 axis than in the galactic plane, though in neither case is the difference especially large. 

\begin{figure}
\includegraphics[width=3in]{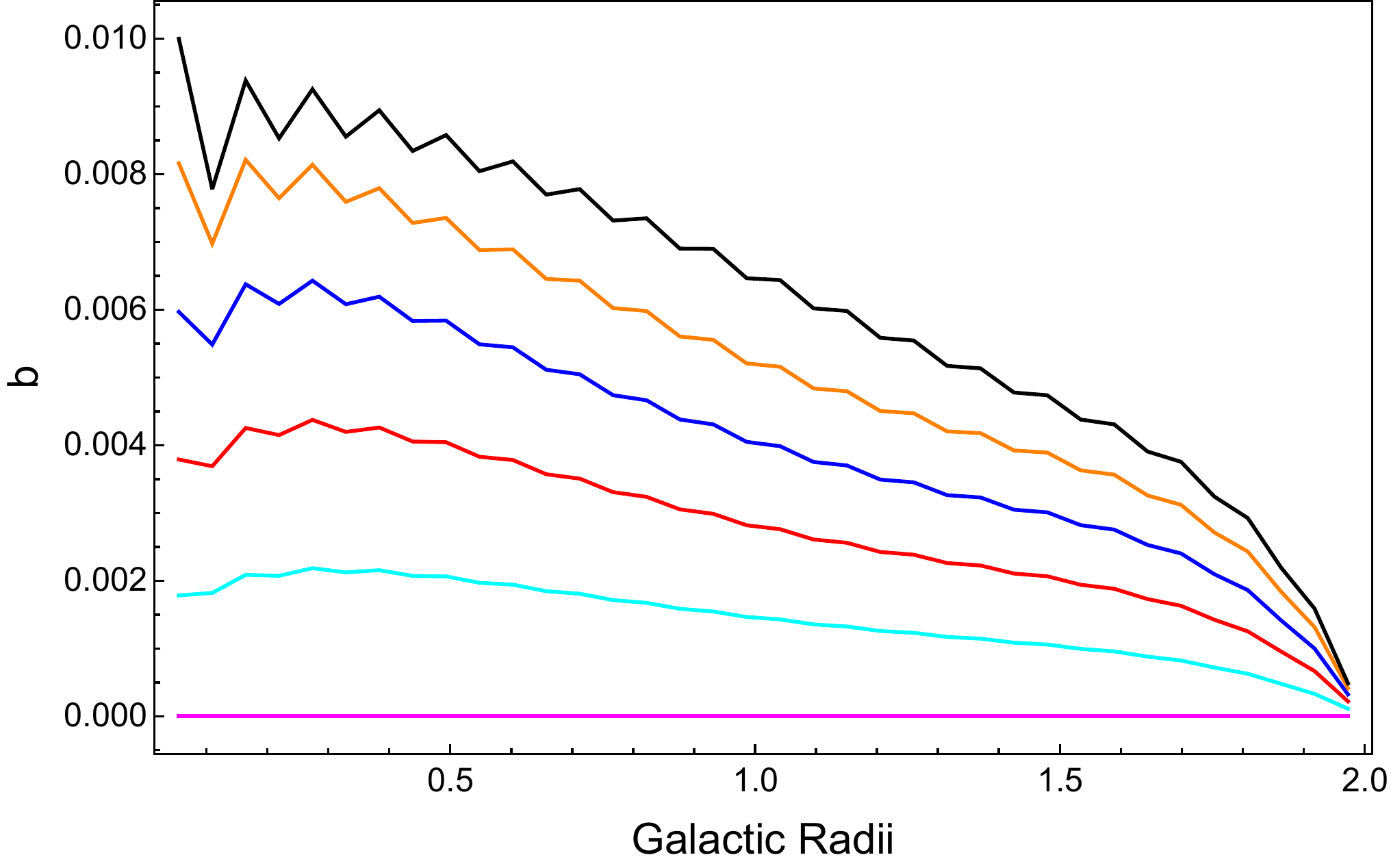}
\caption{The difference between the pristine and quasilinear MOND fields on the 
for various aspect ratios ($b_r$ vs $r$). The vertical axis is in units of $a_0$; the horizontal axis in units of 
galactic radius $R_d$, and mass parameter $\mu = 0.1$. 
The six different curves correspond to $\Delta = 1.0$ (magenta), 0.9 (cyan), 0.8 (red),
0.7 (blue), 0.6 (orange) and 0.5 (black).  As expected the difference between the pristine and the quasilinear fields is zero for an isotropic galaxy.}
\label{fig:b}
\end{figure}

Fig \ref{fig:b} shows the radial component of ${\mathbf b}$ the difference between the
pristine and quasilinear fields plotted as a function of $r$. The plots are for a galaxy with mass parameter $\mu = 0.1$. 
Because the difference of the pristine and quasilinear fields is a small quantity
the finiteness of the grid is visible in the jagged appearance of the plots; the deviations from a smooth fit are an
estimate of the precision of the numerics. 
Another feature lost to low resolution is that close to the
origin the radial component of ${\mathbf b}$ turns down and approaches zero. We have verified that with
higher resolution grids this behavior is seen in the numerics. Furthermore except very near to the origin the
high resolution numerics agree with the coarser plot shown in fig \ref{fig:b}. As one might expect the 
difference between the pristine and quasilinear fields vanishes in the isotropic limit $\Delta = 1$. As 
the disk becomes flatter with decreasing $\Delta$ the difference between the fields grows. The effect
for $\Delta = 0.1$ (the approximate aspect ratio of the Milky Way, for example) is not very different from that of $\Delta = 0.5$.
Since the latter aspect ratio can be simulated accurately with a coarser grid this is the canonical 
aspect ratio we have used in many of our plots. For the purposes of this paper a more 
faithful model of the galaxies is not necessary. It is also interesting to study how the magnitude of
${\mathbf b}$ varies with mass parameter $\mu$ for a fixed aspect ratio $\Delta$. We find that
the magnitude is bigger for a galaxy deep in the MOND regime than for a Newtonian galaxy; in other
words it grows as $\mu$ decreases. 

\begin{figure}
\includegraphics[width=3in]{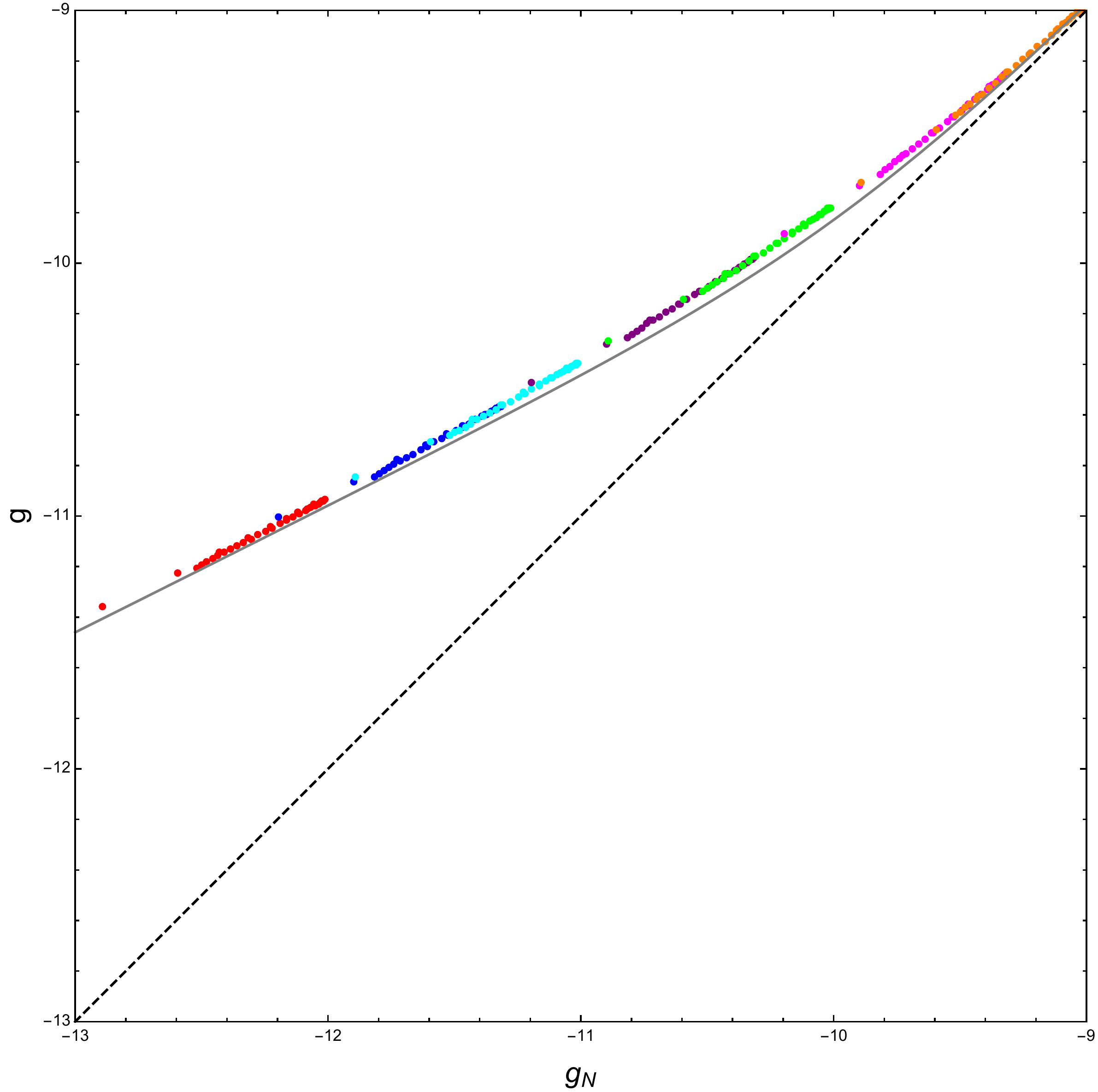}
\caption{Simulated radial acceleration relation for disk galaxies within quasilinear MOND. 
The data points correspond to the radial component of the quasilinear MOND acceleration plotted as a function of
corresponding radial component of the Newtonian acceleration at the same point. For each galaxy, 
the accelerations are calculated at 36 points in the galactic plane at different distances 
from galactic center ranging from zero to two galactic radii. The accelerations are converted to 
units of m/s$^2$ assuming the value $a_0 = 1.2 \times 10^{-10}$ m/s$^2$ and are plotted on 
a logarithmic scale. All galaxies have
$\Delta = 0.5$; mass parameters are $\mu = 0.01$ (red), 0.05 (blue), 0.1 (cyan), 0.5 (purple),
1.0 (green), 5.0 (magenta) and 10.0 (orange). The grey curve shows the radial component of the pristine
MOND acceleration plotted as a function of the Newtonian acceleration; the dashed line is the large field asymptote
wherein the MOND accelerations equal the Newtonian. The deviations of the quasilinear MOND data from the 
pristine MOND curve confirms that the radial acceleration relation has intrinsic scatter in quasilinear MOND. 
However the scatter is small compared to observations uncertainties showing that the observed radial 
acceleration relation is compatible with quasilinear MOND. }
\label{fig:streaky}
\end{figure}

In fig \ref{fig:streaky} we plot the radial component of the pristine and quasilinear MOND acceleration as a 
function of the corresponding Newtonian acceleration for our model galaxies. In comparing this plot
to the radial acceleration relation of ref \cite{stacy1} we note that the Newtonian acceleration corresponds
to the computed baryonic acceleration in ref \cite{stacy1} while the pristine and quasilinear accelerations
correspond to the observed acceleration of ref \cite{stacy1} within pristine and quasilinear MOND theories
respectively. For pristine MOND
by definition the observed acceleration is a simple algebraic function of the baryonic acceleration (grey curve
in fig \ref{fig:streaky}). For quasilinear MOND for our model galaxies there should be a distinct curve for 
each value of the galaxy parameters $\Delta$ and $\mu$. This is indeed borne out by fig \ref{fig:streaky} 
in which we have plotted the quasilinear predictions only for $\Delta = 0.5$ but for seven different values of
$\mu$. However because of the small difference between the pristine and quasilinear accelerations, all of
these curves lie close to the pristine curve, and hence we arrive at our principal conclusion that quasilinear
MOND is compatible with the observed radial acceleration relation. The deviations from the smooth 
pristine behavior are bigger for $\Delta = 0.1$ \cite{onesided}
but even in this case they remain smaller than the
observational uncertainties described in ref \cite{stacy1}. Whether there are systematic trends buried in the
observed deviations when the data are grouped according to parameters like $\mu$ and $\Delta$, trends
that can be predicted by quasilinear MOND, is a difficult problem we leave open for future work.

\begin{figure}
\includegraphics[width=3in]{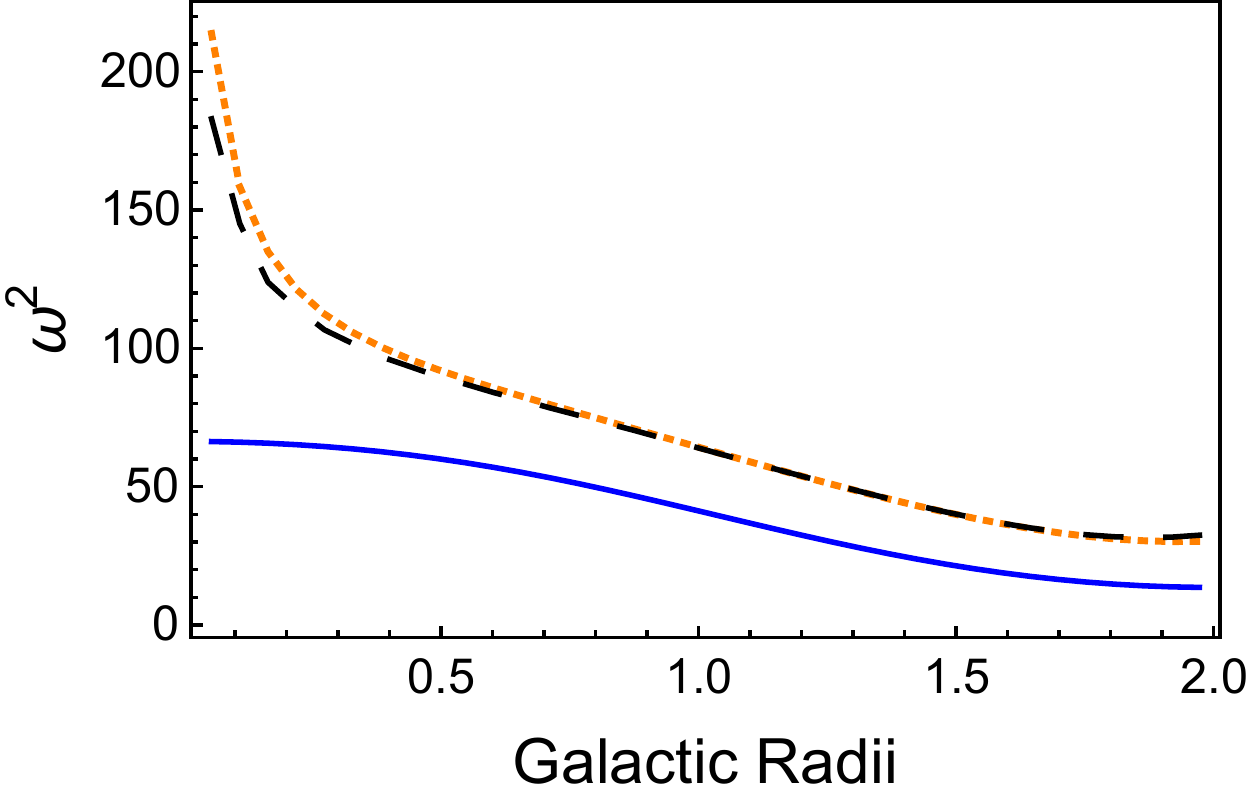}
\caption{Frequency of oscillations perpendicular to the plane for 
particle in a circular orbit for  
quasilinear MOND (dashed), pristine MOND (orange), and Newtonian (blue) fields, assuming $\Delta = 0.5$ and $\mu = 1.$ }
\label{fig:osc}
\end{figure}

Although not directly related to the radial acceleration relation, we now consider oscillatory motion
perpendicular to the galactic plane. A test particle moving in a circular orbit in the galactic plane will
undergo such oscillations if perturbed suitably. If this oscillation frequency could be measured it would provide a clear
distinction between MOND and $\Lambda$CDM. To calculate this frequency we note that in the galactic plane,
by symmetry the $z$ component of the acceleration due to gravity must vanish. Hence near the galactic plane,
the acceleration due to gravity varies linearly with $z$ and hence a particle perturbed away from the galactic
plane executes simple harmonic motion. The frequency of oscillations is given by 
$\omega^2 = - \partial g_z/\partial z$. Here $g_z$ denotes the $z$ component of ${\mathbf g}_Q$ within
quasilinear MOND and of ${\mathbf g}_P$ in the pristine approximation to it. In the $\Lambda$CDM paradigm
$g_z$ denotes the $z$ component of the Newtonian field of visible matter and the dark matter halo combined.
In fig \ref{fig:osc} we show the oscillation frequency computed within quasilinear MOND, the pristine approximation
and due to the Newtonian field of visible matter for a disk galaxy with aspect ratio $\Delta = 0.5$ and mass parameter
$\mu = 1$. The Newtonian field of visible matter dominates the 
$\Lambda$CDM contribution because it constitutes a disk whereas the dark matter is dispersed into a spherical
halo. Taking the visible Newtonian frequency to be an estimate of the total $\Lambda$CDM oscillation frequency
we see that MOND predicts a much higher frequency of oscillation than does $\Lambda$CDM.
The idea of using vertical motion to distinguish MOND from $\Lambda$CDM has been extensively investigated in the literature. Recently Angus et al have fit both rotation curves and vertical velocity dispersion for a sample of galaxies from the DiskMass survey \cite{angus}. Margalit and Shaviv \cite{margalit} have argued that in MOND gravity stars undergoing vertical oscillation drift in their orbits in proportion to the square of the amplitude of their vertical oscillation; for simplicity in their analysis they approximate the quasilinear MOND field by the pristine field, an approximation that is justified by the results depicted in Fig.\ref{fig:osc}. Earlier Bienayme et al \cite{bienayme} and Nipoti et al \cite{nipoti} have proposed  Milky Way tests to distinguish nonlinear MOND predictions from $\Lambda$CDM. Recently models of dark matter have appeared that posit the formation of a disk of dark matter; these models also predict enhanced vertical oscillation frequencies \cite{randall1,randall2}. It is likely that future surveys providing data on both in plane and vertical motion will be needed to discriminate among these different models.

%

\section{Conclusion}

\label{conclusion}

The radial acceleration relation is a plot of the observed radial acceleration in disk galaxies as 
a function of the acceleration that would be expected on the basis of the Newtonian gravitational
field of the visible baryonic matter. Pristine MOND predicts that the relation is a smooth curve
determined by the interpolating function $f$ of the theory. Quasilinear MOND however predicts
a scatter about the smooth curve resulting from Pristine MOND. The principal finding of this paper
is that quasilinear MOND is nonetheless compatible with the observed radial acceleration 
relation because the predicted intrinsic scatter is small compared to the observational uncertainties. 
It appears a daunting task to try to relate the observed scatter about a smooth fit to the 
predicted intrinsic scatter. In principle one could imagine examining the deviations after sorting the
galaxies in to groups according to parameters such as $\mu$ and $\Delta$. However as already
noted \cite{onesided} 
the predicted deviations are even smaller than apparent from the plot in fig \ref{fig:streaky} 
and are moreover model dependent being sensitive to the unknown interpolating function $f$.


\appendix

\section{Newtonian field near galactic center}

\label{sec:appendixa}

Near the origin the Newtonian potential corresponding to the distribution 
eq (\ref{eq:profile}) can be expanded in a power series. Here we show
the form of this series and a sequence of sum rules that the coefficients in the expansion must satisfy.
By verifying that the sum rules are satisfied we are able to perform a non-trivial check
on our numerical computations. 

The Newtonian potential is related to the Newtonian field via ${\mathbf g}_N = - \nabla \phi_N$.
We take $\phi_N = 0$ at the origin. Taking into account the azimuthal and reflection symmetries
of the galaxy profile (\ref{eq:profile}), to quartic order the potential must have the form
\begin{equation}
\phi_N = A (x^2 + y^2) + B z^2 + C (x^2 + y^2)^2 + D (x^2 + y^2) z^2 + E z^4 + \ldots
\label{eq:expansion}
\end{equation}
It is now a straightforward matter to compute the Laplacian of the Newtonian potential.
Comparing to the corresponding series expansion of the density $\rho_D$ and making use
of Poisson's equation for Newtonian gravity, $\nabla^2 \phi_N = 4 \pi G \rho$, we deduce
\begin{eqnarray}
4 A + 2 B & = & 4 \mu /\Delta \sqrt{\pi}, \nonumber \\
16 C + 2 D & = & - 4 \mu / \Delta \sqrt{\pi}, \nonumber \\
4 D + 12 E & = & - 4 \mu / \Delta^3 \sqrt{\pi}.
\label{eq:sumrules}
\end{eqnarray}

Numerically the coefficients $A, B, C, D$ and $E$ are computed by differentiating the 
Newtonian field ${\mathbf g}_N$ at the origin. We compute these derivatives in Fourier space.
The numerically calculated coefficients satisfy the sum rules to a part in a thousand accuracy.
The agreement confirms not only that our calculation of the Newtonian field is accurate near the origin
but also that our scheme of computing space derivatives in Fourier space is accurate. This is
reassuring since we do need to compute spatial derivatives in order to compute quantities of interest
such as e.g. $\nabla \cdot {\mathbf g}_P$ and $\nabla \times {\mathbf g}_P$.


\begin{thebibliography}{999}
\bibitem{olbers} 
The argument is similar to the reasoning
that leads to Olbers paradox in cosmology. An analogous difficulty also arises in solid state physics in calculating
the energy of ionic crystals and is resolved in a similar fashion.
\bibitem{onesided}
The deviations from the smooth curve in fig \ref{fig:streaky} are all
positive because both the quasilinear data and the smooth pristine curve are based on the same 
interpolating function with the same value for the MOND parameter $a_0$. If the MOND parameter
were treated as a fitting parameter an even better smooth fit could be found for the same quasilinear MOND data.
\end{thebibliography}
\end{document}